\documentstyle[preprint,aps,amsmath,amsfonts,latexsym,amssymb,bbold,graphicx,rotating]{revtex}

\begin{document}

\title{Winding Number Transitions in the Mottola--Wipf Model on a Circle}
\author{D.K. Park$^{1,2}$\footnote{e-mail:
dkpark@genphys.kyungnam.ac.kr},
H. J. W. M\"{u}ller-Kirsten$^{1}$\footnote{e-mail:
mueller1@physik.uni-kl.de}and J.--Q.
Liang$^{1,3}$\footnote{e-mail:jqliang@mail.sxu.edu.cn}}
\address{1. Department of Physics, University of Kaiserslautern, D-67653 Kaiserslautern, Germany\\
2.Department of Physics, Kyungnam University, Masan, 631-701, Korea\\
3. Department of Physics and Institute of Theoretical Physics,
Shanxi University, Taiyuan, Shanxi 030006, China}
\maketitle

\begin{abstract}
Winding number transitions from quantum to classical behavior
are studied in the case of the {1+1} dimensional Mottola--Wipf
model with the space coordinate on a circle for exploring
the possibility of obtaining transitions of second order. The model
is also studied as a prototype theory which demonstrates
the procedure of such investigations. In the model at hand we
find that even on a circle the transitions remain those of first order. 
\end{abstract}

%\centerline{PACS numbers:75.45.+j, 75.50.Tt} 

\section{Introduction}
\label{sec:I}
The study of transitions from quantum behavior to classical behavior
has attracted considerable attention recently as a result
of extension of semiclassical vacuum instanton considerations to
those of periodic instantons and sphalerons, and the consequent
possibility to study such theories at higher temperatures 
\cite{Chudnovsky,Liang1,Lee1}. In these
earlier investigations the order of the
transitions was inferred from monotonically decreasing or nondecreasing
behavior of the period of the periodic pseudoparticle configurations.
But in other contexts \cite{Gorokhov}it was shown
that one can derive criteria for the occurrence of such phase
transitions of one type or the other by expanding the field
concerned about the sphaleron configuration. This idea
was then further exploited \cite{Rana}to obtain concrete conditions
in the form of inequalities
for a large class of quantum mechanical models, applications of which
can be found in \cite{Park1,Zhang1,Park2,Park3,Park4}.  

The period $\beta$ is related to the energy $E$
in the standard way, i.e.  $E=\frac{\partial S}{\partial\beta}$,
where $S(\beta )$ is the action of the periodic
instanton (bounce) per period. Such
periodic instantons (bounces) smoothly interpolate
between the  zero temperature instantons (bounces)
and the static solution named sphaleron at
the top of the potential barrier. The sphaleron
is  responsible for thermal hopping.
With increasing temperature thermal hopping becomes more and more
important and beyond some critical or crossover temperature $T_c$
becomes the decisive mechanism.
It is more difficult to study such phase transitions in 
the context of field theory. However,
as mentioned, the criterion for a first
order transition can be obtained by studying the Euclidean
time period in the neighborhood of the sphaleron.
If the period $\beta (E\rightarrow U_{0})$ of the periodic instanton
(bounce) close to the barrier peak can be found,
a sufficient condition for a
first order transition is seen to be the inequality
$\beta (E\rightarrow U_{0})-\beta_{s} 
< 0$ or $\omega^{2} > \omega^{2}_{s}$,
where $U_{0}$ denotes the barrier height and $\beta_{s}$ is
the period of small oscillation around
the sphaleron. Here $\omega $ and $ \omega _{s}$ are
the corresponding frequencies.
The frequency of the sphaleron $\omega _{s}$ is
the frequency of small
oscillatons at the bottom of the inverted potential well.
The winding number transition
in the $O(3) \sigma$ model with and without
a Skyrme term has been successfully
analysed with such a criterion in ref.\cite{Park1};
other applications have been carried out in refs.
\cite{Zhang1,Park2,Park3,Park4,Habib}

In ref.\cite{Park4} the theory defined by the following  euclidean
action was considered
\begin{eqnarray}
S_E&=&\int d\tau dx\bigg[\frac{1}{2}\bigg(\frac{d\phi}{d\tau}\bigg)^2
+ \frac{1}{2}\bigg(\frac{d\phi}{dx}\bigg)^2+U(\phi)\bigg],\nonumber\\
U(\phi)&=&-\frac{\mu^2}{2a^2}(\phi^2-a^2)^2+\frac{\mu^2}{2a^2}a^4.
\label{1}
\end{eqnarray}
It was found that the quantum classical transition is always of the
smooth second order type for a noncompactified spatial coordinate. If
one considers the same theory with $x$ on $S^1$, the theory
can become first order when the elliptic modulus $k$ of the
periodic solution is subjected to a condition $k<k^*<1$.  The limit
of decompactification (i.e. $S^1\rightarrow\mathbb{R}$) is given by $k=1$.
If we write our criterion inequality $f<0$ and plot $f$ versus $k$,
then a first--order transition implies that $f<0$ for $k<k^*$.
Thus one can conclude that either (i) compactification prefers
a first--order transition, or (ii) compactification changes the
type of transition. In the following we investigate which of these
possibilities is right in the case of the compactified 
Mottola--Wipf model, for which the uncompactified limit is
always of first--order, as shown in \cite{Park1,Habib}.
Since the Mottola--Wipf model serves as an important testing ground
of many field theoretical
aspects, it is desirable to explore what happens in
its case with compactification.
If (i) is
correct, the compactified case implies a stronger tendency to
first--order behavior, if (ii) is correct one expects a sign
change in $f$, and so a rising behavior below a critical value of  
$k$. In the following we shall see that case (i) applies.

\section{The sphaleron and expansion around it}
\label{sec:II}
The Euclidean time action of the Mottola--Wipf model \cite{MW} is given by
\begin{equation}
S=\frac{1}{g^2}\int d\tau dx
\bigg[\frac{1}{2}\partial_{\mu}n_a\partial_{\mu}n_a+(1+n_3)\bigg]
\label{2}
\end{equation}
where $a=1,2,3, n_a n_a=1,\mu=\tau,x $. The equation of motion is 
\begin{equation} 
(\delta_{ab}-n_an_b)\Box n_b=(\delta_{a3}-n_an_3).
\label{3}
\end{equation}
The static and so only $x$--dependent sphaleron solution is given by
\begin{equation} 
{\bf n}_{sph}(x)=(-\sin f_s(x), 0, \cos f_s(x)),\;\;\;f_s(x)=2\arcsin[k sn(x)]
\label{4}
\end{equation}
where $k$ is the elliptic modulus of the Jacobian elliptic
functions involved and (later) ${k^{\prime}}^2=1-k^2$.
For the expansion around the sphaleron configuration we set
\begin{equation}
 {\bf n}(x,\tau)=\frac{1}{\sqrt{1+u^2}}\bigg(-\sin(f_s+v), u, \cos(f_s+v)\bigg)
\label{5}
\end{equation}
where $u(x,\tau)$ and $v(x, \tau)$ are fluctuation fields. Inserting
this ansatz into the equation of motion, we obtain after some
tedious calculations the following set of equations
\begin{equation}
{\hat{\vartheta}}{u\choose v}={\hat{h}}
{u\choose v}+{{G^u_2(u,v)}\choose {G^v_2(u,v)}}+
{{G^u_3(u,v)}\choose {G^v_3(u,v)}}+\cdot\cdot\cdot
\label{6}
\end{equation}
with operators
\begin{equation}
{\hat{\vartheta}}=\left(\begin{array}{cc}
\displaystyle \frac{\partial^2}{\partial\tau^2}&\displaystyle 0\\
0 & \displaystyle \frac{\partial^2}{\partial\tau^2}\end{array}\right),\;\;\;
\hat{h}=\bigg(\begin{array}{cc}
\displaystyle \hat{h}_u&\displaystyle 0\\
0 & \displaystyle \hat{h}_v\end{array}\bigg),
\label{7}
\end{equation}
In these equations
\begin{eqnarray}
G^u_2(u,v)&=&2ksn(x) dn(x)uv- 4kcn(x) uv^{\prime}\nonumber\\
G^v_2(u,v)&=&-ksn(x) dn(x) (u^2-v^2) + 4kcn(x) uu^{\prime}\nonumber\\
G^u_3(u,v)&=&2u(\dot{u}^2+{u^{\prime}}^2)-u(\dot{v}^2 + {v^{\prime}}^2)
-\frac{1}{2}(1-2k^2 sn^2(x))u(u^2-v^2)\nonumber\\
G^v_3(u,v)&=&\frac{1}{6}(1-2k^2 sn^2(x))(v^3-3u^2v)+2u({\dot u}{\dot v}+
u^{\prime}v^{\prime})
\label{8}
\end{eqnarray}
where the dot and prime imply differentiation
with respect to $\tau$ and $x$ respectively,  and
\begin{equation}
\hat{h}_u = -\frac{\partial^2}{\partial x^2}-(1+4k^2-6k^2sn^2(x)),\;\;\;
\hat{h}_v=-\frac{\partial^2}{\partial x^2}-(1-2k^2sn^2(x)).
\label{9}
\end{equation}
The equations 
\begin{equation}
\hat{h}_u\psi_n(x)=\epsilon^u_n\psi_n(x),\;\;\;
\hat{h}_v\phi_n(x)=\epsilon^v_n\phi_n(x) 
\label{10}
\end{equation}
are Lam$\acute{e}$ equations. The $2N+1, N=2,1$ discrete eigenvalues of these
equations and their respective eigenfunctions whose
periods are $4K(k)$ and $2K(k)$ (cf.\cite{Arscott})
are given in Table 1.

\newpage
\centerline{\bf Table 1}

\centerline{Discrete eigenfunctions and eigenvalues of $\hat{h}_u, \hat{h}_v$}

\vspace{0.2cm}
\begin{center}
\begin{tabular}{|c|c|c|c|}\hline
 $n$ &  Period & $\psi_n(x)$ & $\epsilon^u_n$\\ \hline\hline
$0$ & $2K$ & $\psi_0(x)=C_0
[sn^2(x)-\frac{1+k^2+\sqrt{1-k^2{k^{\prime}}^2}}{3k^2}]$&
$\epsilon^u_0=1-2k^2-2\sqrt{1-k^2{k^{\prime}}^2}$\\ \hline
$1$ & $4K$ & $\psi_1(x)=C_1cn(x) dn(x)$ & $\epsilon^u_1=-3k^2$\\ \hline
$2$ & $4K$ & $\psi_2(x)=C_2 sn(x) dn(x)$ & $\epsilon^u_2 = 0$\\ \hline 
$3$ & $2K$ & $\psi_3(x) = C_3 sn(x) cn(x)$ & $\epsilon^u_3 = 3(1-k^2)$\\ \hline
$4$ & $2K$ & $\psi_4(x) = C_4[sn^2(x) -
\frac{1+k^2-\sqrt{1-k^2{k^{\prime}}^2}}{3k^2}]$
& $\epsilon^u_4 = 1-2k^2+2\sqrt{1-k^2{k^{\prime}}^2}$\\ \hline\hline
\end{tabular}
\vspace{0.2cm}

\vspace{0.2cm}
\begin{tabular}{|c|c|c|c|}\hline
 $n$ &  Period & $\phi_n(x)$ & $\epsilon^v_n$\\ \hline\hline
$0$ & $2K$ & $\phi_0(x)=D_0 dn(x)$ &
$\epsilon^v_0=-(1-k^2)$\\ \hline
$1$ & $4K$ & $\phi_1(x)=D_1cn(x)$ & $\epsilon^v_1=0$\\ \hline
$2$ & $4K$ & $\phi_2(x)=D_2 sn(x)$ & $\epsilon^v_2 = k^2$ \\ \hline\hline
\end{tabular}
\end{center}

We now consider the spectrum of $\hat{h}$ and write
\begin{equation}
\hat{h}{\bf \xi}_n = h_n{\bf \xi}_n.
\label{11}
\end{equation}
The eigenfunctions and respective eigenvalues of eq.(\ref{11}) depend,
of course, on the boundary conditions of the theory.  We set
$$
{\bf n}(x,\tau)=(n_1(x,\tau),n_2(x,\tau), n_3(x,\tau))
$$
and consider the following boundary conditions:

(i){\it Periodic boundary conditions}

\begin{equation}
n_1(x+4K,\tau)=n_1(x,\tau),n_2(x+4K,\tau)=n_2(x,\tau),
n_3(x+4K,\tau)=n_3(x,\tau)
\label{12}
\end{equation}

(ii){\it Partially anti--periodic boundary conditions}

\begin{equation}
n_1(x+2K,\tau)=-n_1(x,\tau),n_2(x+2K,\tau)=-n_2(x,\tau),
n_3(x+2K,\tau)=n_3(x,\tau)
\label{13}
\end{equation}

Then up to the first order
\begin{equation}
{\bf n}={\bf n}_{sph}+\delta{\bf n}
\label{14}
\end{equation}
where
\begin{equation}
{\bf n}_{sph}=(-2k sn(x) dn(x), 0, 2dn^2(x)-1),\;\;
\delta{\bf n}=(-(2dn^2(x)-1)v, u, -2k sn(x) dn (x) v).
\label{15}
\end{equation}
Thus in the case of periodic boundary conditions we need
fluctuation fields $u$ and $v$ which have the period $4K$
and all 8 cases of Table 1 are possible.
The eigenfunctions and eigenvalues of eq.(\ref{11})
are then of the following
type:
\begin{eqnarray}
{\bf \xi}_0&=&(\psi_0,0),\;\;\;\epsilon^u_0<0,\nonumber\\
{\bf \xi}_1&=&(\psi_1,0),\;\;\;\epsilon^u_1<0,\nonumber\\
{\bf \xi}_2&=&(0,\phi_0),\;\;\;\epsilon^v_0<0,\nonumber\\
{\bf \xi}_3&=&(\psi_2,0),\;\;\;\epsilon^u_2=0,\nonumber\\
{\bf \xi}_4&=&(0,\phi_1),\;\;\;\epsilon^v_1=0,\nonumber\\
{\bf \xi}_5&=&(\psi_3,0),\;\;\;\epsilon^u_3>0,\nonumber\\
{\bf \xi}_6&=&(0,\phi_2),\;\;\;\epsilon^v_2>0,\nonumber\\
{\bf \xi}_7&=&(\psi_4,0),\;\;\;\epsilon^u_4>0.
\label{16}
\end{eqnarray}
Hence ${\hat h}$ has three negative modes and two zero modes in this case.  
Thus it is impossible to analyse the phase transition problem
properly in the case of periodic boundary conditions.

In the second case of partially antiperiodic boundary conditions
our findings of the spectrum of ${\hat h}$ are as follows.
Considering again eq.(\ref{15}) one can show that
with $2K$--antiperiodic $u$ and $v$ satisfying
$$
u(x+2K)=-u(x),\;\;\;v(x+2K)=-v(x)
$$
the eigenfunctions ${\bf \xi}_n$ and eigenvalues $\epsilon_n$ become
those given in Table 2 below.  
\newpage
\centerline{\bf Table 2}

\centerline{Eigenfunctions and eigenvalues of ${\hat h}$ in case of
partially antiperiodic boundary conditions}

\vspace{0.2cm}

\begin{center}
\begin{tabular}{|c|c|c|}\hline
 $n$ &  ${\bf \xi}_n$ & $h_n$\\ \hline\hline
$0$  & ${\bf \xi}_0=(\psi_1,0)$ &$h_0=\epsilon^u_1=-3k^2$
\\ \hline
$1,2$ & ${\bf \xi}_1=(\psi_2,0),{\bf \xi_2}=(0,\phi_1)$ & $h_1=h_2=0$\\ \hline
$3$ & ${\bf \xi}_3=(0,\phi_2)$ & $h_2=\epsilon^v_2=k^2$ \\ \hline\hline
\end{tabular}
\end{center}
Since this case produces only one negative eigenmode we can consider
the winding number transition with our method.
Hence, in the next section we will investigate the winding number 
phase transition in this model with the partially antiperiodic boundary
condition. Although the periodic boundary condition in the model is fully 
examined in Ref. \cite{fun90}, we think the partially antiperiodic boundary
condition is physically more reasonable in the sense that it has an 
uncompactified limit which coincides with the spectrum derived in the 
{\it a priori} infinite spatial domain. This fact justifies our choice
of this boundary condition.

\section{Derivation of the criterion inequality}
\label{sec:III}
The calculation of the inequality proceeds as in \cite{Park1}. 
Without repeating detailed calculational steps we
summarize briefly how the criterion for the sharp first-order transition
is obtained. Let $u_0$ and $\epsilon_0$ be eigenfunction and eigenvalue
of the negative mode of the fluctuation operator. Therefore, the sphaleron
frequency $\Omega_{sph}$ is defined as $\Omega_{sph} \equiv \sqrt{-
\epsilon_0}$. Then the type of the transition is determined by studying 
the nonlinear corrections to the frequency. Let, for example, $\Omega$ be a
frequency involving the nonlinear corrections. If $\Omega_{sph} - \Omega > 0$,
the period-vs-energy diagram has monotonically decreasing behavior at least
in the vicinity of the sphaleron. This means $dS/dT$, where 
$S$ and $T$ are classical 
instanton action and temperature(inverse of period), is monotonically
decreasing around the sphaleron and eventually, $S$ merges with 
the sphaleron
action smoothly as shown in Ref. \cite{Chudnovsky}. If, on the contrary,
$\Omega_{sph} - \Omega < 0$, the temperature dependence of instanton 
action consists of monotonically decreasing and increasing parts, which
results in the discontinuity of $dS/dT$. This is the main idea which is used
to derive the criterion in Ref. \cite{Park1}.

Recently, similar considerations have been used in the context of 
the $SU(2)$-Higgs model
\cite{fro99,bon99}, where the winding number transition is 
characterized by the monotonic sign of $d^2S/d\beta^2$, where $\beta$ is the
period.
From the viewpoint of our criterion it is clear that the non-monotonic
behavior of $dS/d\beta$ exactly corresponds to $\Omega_{sph} - 
\Omega < 0$. 

Now, we will compute $\Omega$ following the procedure in Ref. \cite{Park1}.
We
restrict ourselves to a presentation of the main steps. We set
(cf. Table 1)
$$
u_0(x)=\psi_1(x)=C_1 cn(x) dn(x)
$$
Using $\int^{K}_{-K} dx \psi^*_1(x)\psi_1(x)=1$,
the normalization constant $C_1$ is
found to be
\begin{equation}
C_1=\bigg[\frac{2}{3k^2}\bigg\{(1+k^2)E-(1-k^2)K\bigg\}\bigg]^{-\frac{1}{2}}
\label{17}
\end{equation}
where $K(k)$ is the quarter period introduced earlier and $E(k)$
the complete elliptic integral of the second kind
(here and elsewhere we use formulas of ref.\cite{Byrd}).
We let $a$ be a small amplitude around the sphaleron.  
Then, one can carry out the perturbation with an expansion parameter $a$.
In the first order perturbation where we have to expand up to the quadratic
terms in Eq.(\ref{6}) it is shown that $\Omega$ does not have a correction.
The next order perturbation where we have to consider terms 
up to the cubic terms
generates a nonlinear correction to $\Omega$, which makes the condition for a 
first order transition to be 
\begin{equation}
{\Omega^2_{sph}}-{\Omega}^2 = a^2<u_0|G_{u,1}> \;\;\;<0
\label{18}
\end{equation}
where $<\psi_1|\psi_2> \equiv \int_{-K}^{K} dx \psi_1^*(x) \psi_2(x)$, 
$\Omega_{sph} = \sqrt{3} k$,  
\begin{eqnarray}
G_{u,1}(x)&=&2k sn(x) cn(x) dn^2(x)(g_{v,1}(x)+g_{v,2}(x)/2)
-4kC_1 cn^2(x) dn(x)(g^{\prime}_{v,1}(x)+g^{\prime}_{v,2}(x)/2)\nonumber\\
&+&\frac{C_1^3}{4}\bigg[cn^3(x) dn^3(x)
\{2\Omega^2_{sph}-\frac{3}{2}+3k^2 sn^2(x)\}\nonumber\\
&+&6sn^2(x) cn(x)dn(x)\{dn^2(x)+k^2cn^2(x)\}^2\bigg]
\label{19}
\end{eqnarray}
and here
\begin{equation}
g_{v,1}(x)=\frac{k}{2}C^2_1{\hat {h_v}}^{-1}
\bigg[\{(5+4k^2)-9k^2sn^2(x)\}sn(x) cn^2(x) dn(x)\bigg]
\label{20}
\end{equation}
and
\begin{equation}
g_{v,2}(x)=\frac{k}{2}C^2_1({\hat {h_v}}+4\Omega^2_{sph})^{-1}
\bigg[\{(5+4k^2)-9k^2sn^2(x)\}sn(x) cn^2(x) dn(x)\bigg].
\label{21}
\end{equation}
The inequality (\ref{18}) can be written in the form
\begin{equation}
\frac{{\Omega^2_{sph}}-{\Omega}^2}{a^2}=I_1(k)+I_2(k)+I_3(k)+I_4(k) < 0
\label{22}
\end{equation}
where
\begin{equation}
I_i(k)=<u_0|f_i>, \;\;\; i=1,2,3,4,
\label{23}
\end{equation}
with
\begin{eqnarray}
f_1&=&2kC_1sn(x)cn(x)dn^2(x)g_{v,1}-4kC_1cn^2(x)dn(x)g^{\prime}_{v,1}\nonumber\\
f_2&=&kC_1sn(x)cn(x)dn^2(x)g_{v,2}-2kC_1cn^2(x)dn(x)g^{\prime}_{v.2}\nonumber\\
f_3&=&\frac{C_1^3}{4}cn^3(x)dn^3(x)
\{2{\Omega^2_{sph}}-\frac{3}{2}+3k^2sn^2(x)\}\nonumber\\
f_4&=&\frac{3}{2}C^3_1sn^2(x)cn(x)dn(x)\{dn^2(x)+k^2cn^2(x)\}^2.
\label{24}
\end{eqnarray}

\section{Evaluation of the quantities}{\vspace{-1.35cm}\hspace{12cm} $I_i(k)$}
\label{sec:IV}

It is now necessary to evaluate the quantities entering the
criterion inequality. Since this is nontrivial
we consider these individually.

\centerline{\bf The quantity $I_1(k)$}

We consider first the case of $k=1$. 
In this case $g_{v,1}(x)$ becomes
\begin{equation}
g_{v,1}(k=1,x)=\frac{27}{8}{\hat {h_v}}\frac{\sinh x}{\cosh^6 x},
\;\;\;{\hat {h_v}}^{-1}(k=1)=-\frac{\partial^2}{{\partial x}^2}+(1-2/\cosh^2 x).
\label{25}
\end{equation}

Since the operator ${\hat {h_v}}^{-1}(k=1)$ is of the
usual P\"oschl--Teller type,
the complete spectrum can be obtained and is summarised in
Table 3.  

\centerline{\bf Table 3}

\centerline{Eigenfunctions and eigenvalues of ${\hat h}^{-1}_v(k=1)$}

\vspace{0.2cm}

\begin{center}
\begin{tabular}{|c|c|}\hline
Eigenvalue of ${\hat {h_v}}^{-1}(k=1)$ &  Eigenfunction
 of  ${\hat {h_v}}^{-1}(k=1)$\\ \hline\hline
Discrete mode: 0 & $<x|0>=\frac{1}{\sqrt{2}}\frac{1}{\cosh x}$
\\ \hline
Continuum mode:$1+k^2_c $ & $ <x|k_c>=\frac{e^{ik_cx}}
{\sqrt{2\pi}(1+ik_c)}(ik_c-\tanh x)$ \\ \hline\hline
\end{tabular}
\end{center}

Using the completeness relation one can evaluate 
$g_{v,1}(k=1,x)$. We find
\begin{eqnarray}
g_{v,1}(k=1,x)&=&<x|\frac{27}{8}{\hat h}^{-1}_v(k=1)\bigg[|0><0|+
\int dk_c|k_c><k_c|\bigg]|x>\frac{\sinh x}{\cosh^6x}\nonumber\\
&=&\frac{3}{16}\frac{\sinh x}{\cosh^4x}(1+2\cosh^2x).
\label{26}
\end{eqnarray}
By inserting this into eq.(\ref{25}) one can verify that this is
correct. The $k=1$ limit of $I_1(k)$ can now be easily obtained as
\begin{equation}
I_1(k=1)=-\frac{36}{35}=-1.02857.
\label{27}
\end{equation}
Since this is negative it is indicative of support for
a first order contribution.

In order to obtain the k--dependence of $I_1(k)$, we can proceed in
two ways: We can derive first an approximate expression, and then the
exact one. It is instructive to derive the approximation first.
Thus from eq.(\ref{20})  
\begin{eqnarray}
g_{v,1}&=&\frac{k}{2}C^2_1{\hat {h_v}}^{-1}
\bigg[\{(5+4k^2)-9k^2sn^2(x)\}sn(x) cn^2(x) dn(x)\bigg]\nonumber\\
&=&\frac{k}{2}C^2_1\sum_n\frac{1}
{\epsilon^v_n}<\phi_n|\{(5+4k^2)-9k^2sn^2(x)\}
sn(x)cn^2(x)dn(x)>|\phi_n>.
\label{28}
\end{eqnarray}
Since we know only $|\phi_1>$ and $|\phi_2>$
(the former, the zero mode, does not
contribute to (\ref{28})), we approximate the expression
(\ref{28}) as follows.
\begin{eqnarray}
g_{v,1}(k,x)
&=&\frac{k}{2}C^2_1\frac{1}{\epsilon^v_2}<\phi_2|[(5+4k^2)-9k^2sn^2(x)]sn(x)
cn^2(x)dn(x)|\phi_2>\nonumber\\
&=&\frac{\pi C^2_1D^2_2}{32k}(10-k^2)sn(x)
\label{29}
\end{eqnarray}
where
\begin{equation}
D_2=\frac{1}{\sqrt{\frac{2}{k^2}(K-E)}},
\label{30}
\end{equation}
$K$ being the quarter period introduced earlier and
$E$ the complete elliptic integral of the second kind. 
The approximation is valid provided $|\phi_2>$ is an isolated
discrete mode and the density of higher states is dilute.
But if one checks the spectra of ${\hat h}_v(k=1)$ (see also below) and
${\hat h}_v$ in general,
one can see that $|\phi_2>$ is the lowest continuum mode
of  ${\hat h}_v(k=1)$. This is why $D_2(k=1)=0$, and so this approximation
is not valid around $k=1$, as can also be seen from a plot
of $I_1(k)$ with this approximation in Fig.1.

In Appendix A we derive the exact value of $g_{v,1}(k,x)$ using the
zero mode of ${\hat h}_v$.  The result is
\begin{eqnarray}
g_{v,1}(k,x)=\frac{k}{2}C^2_1cn(x)&\bigg[&
\frac{1+k^2(2-4K(k))-3k^4}{4k^2(1-k^2)}E(x)+\frac{4k^2R(k)+k^4-1}{4k^2}x
\nonumber\\
&+&\frac{1}{2}sn(x)cn(x)dn(x)+\frac{R(k)}{1-k^2}\frac{sn(x)dn(x)}{cn(x)}
\bigg]
\label{31}
\end{eqnarray}
where
\begin{eqnarray}
R(k)&=&\frac{1-k^2}{4k^2}\frac{(1+3k^2)E-(1-k^2)(1+k^2)K}{E-(1-k^2)K}
\nonumber\\
E(x)&\equiv &\int^{am x}_0\sqrt{1-k^2\sin^2\theta}
d\theta\equiv\int^x_0dn^2(u)du.
\label{32}
\end{eqnarray}
In the limit $k\rightarrow 1$ these quantities are such that
\begin{equation}
\bigg\{R(k)\bigg\}_{\stackrel{lim}{k\rightarrow 1} }= 0, \;\;\;
\bigg\{\frac{R(k)}{1-k^2}\bigg\}_{\stackrel{lim}{k\rightarrow 1}} =1 
\label{33}
\end{equation}
and
\begin{equation}
g_{v,1}(k=1,x)=\frac{3}{16}\frac{\sinh x}{\cosh^4x}\bigg[1+2\cosh^2x\bigg]
\label{34}
\end{equation}
in agreement with eq.(\ref{26}).
The expression $I_1$ can now be evaluated numerically; its behavior
is also shown in Fig.1. One can see agreement with our $k\rightarrow 1$ 
limit, and that our initial approximation applies in the small $k$
region. From the latter we deduce that the Lam$\acute{e}$
equation has very dilute
discrete higher states in that domain.

\centerline{\bf The quantity $I_2(k)$}

Using the spectrum of ${\hat h}_v(k=1)$ one can derive an
integral representation of $g_{v,2}(k=1,x)$, i.e.
\begin{eqnarray}
g_{v,2}(k=1,x)&=&\frac{3}{256}\int dk_c
\frac{(k^2_c+1)(k^2_c+9)}
{(k^2_c+13)\cosh{\frac{k_c\pi}{2}}}\sin k_cx\nonumber\\
&+&\tanh x\int dk_c \frac{(k^2_c+1)(k^2_c+9)}
{(k^2_c+13)\cosh{\frac{k_c\pi}{2}}}\cos k_cx.
\label{35}
\end{eqnarray}
With this we obtain for the $k=1$ limit of $I_2(k)$
\begin{equation}
I_2(k=1)=-\frac{9\pi}{8192}K_c\approx -0.09683, \;\;\; 
K_c\equiv \int^{\infty}_0\frac{(k^2_c+1)^3(k^2_c+9)^2}{(k^2_c+13)
\cosh^2\frac{k_c\pi}{2}}
\approx 28.0549.
\label{36}
\end{equation}
The negative sign of $I_2$ indicates that it also contributes to
make the transition of first order.
To our knowledge an exact evaluation of $I_2(k)$ is not possible. 
We therefore adopt for two
reasons the above approximate procedure used for the calculation
of $I_{1}(k)$.
(i) From the result of the limit $k=1$ one can conjecture
that $I_2(k)$ is very small in magntitude compared with $I_1(k)$; 
hence its contribution would not change the type of transition.
(ii) From a knowledge of the type of transition at $k=1$, the
interest is shifted to the domain of small $k$, and in this region
our approximate procedures are valid.
In this approximation
\begin{equation}
g_{v,2}(k,x)\approx \frac{\pi C^2_1 D^2_2}{416k}(10-k^2)sn x
\label{37}
\end{equation}
and
\begin{equation}
I_2(k) \approx - \frac{\pi^2C^4_1D^2_2}{6656}(10-k^2)^2.
\label{38}
\end{equation}
The behavior of $I_2(k)$ as a function of $k$ is shown in Fig. 2.

\centerline{\bf The quantity $I_3(k)$}

In the case of $I_3(k)$ we obtain
\begin{eqnarray}
I_3(k)&=&\frac{C^4_1}{420k^4}\bigg[(1-k^2)(-2+111k^2-297k^4-44k^6)K\nonumber\\
&+&(2-112k^2+234k^4+428k^6-88k^8)E\bigg]
\label{39}
\end{eqnarray}
For $k=1$ the value is given in ref.\cite{Park1}:
\begin{equation}
I_3(k=1)=\frac{87}{140}\approx 0.621429.
\label{40}
\end{equation}
Since this is positive, it would help towards a second order transition.
With these results one can show that $I_3(k)$ has the correct $k=1$ limit. 
The behavior of $I_3(k)$ as a function of $k$ is shown in Fig.3.  

\centerline{\bf The quantity $I_4(k)$}

Proceeding as in the other cases the final result is found to be
\begin{eqnarray}
I_4(k)&=&\frac{3C^4_1}{315k^4}\bigg[-(1-k^2)(10-21k^2+48k^4-5k^6)K
\nonumber\\
&+&(10-26k^2+96k^4-26k^6+10k^8)E\bigg].
\label{41}
\end{eqnarray}
At $k=1$ this is
\begin{equation}
I_4(k=1)=\frac{12}{35}\approx 0.342857
\label{42}
\end{equation}
which also supports a tendency towards a second order transition. One
can show that $I_4(k=1)$ has the correct $k=1$ limit.
The behavior of  $I_4(k)$ as a function of $k$ is shown in Fig.4.

\section{Summary and conclusions}
\label{sec:V}
Summarising the above results in the limit $k\rightarrow 1$, we
obtain
\begin{eqnarray}
\frac{{\Omega^2_{sph}}-{\Omega}^2}{a^2}&=&
\sum^4_{i=1}I_i(k=1)=-\frac{36}{35}-\frac{9\pi}{8192}K_c+\frac{87}{140}
+\frac{12}{35}\nonumber\\
&=&-1.02857-0.09683+0.62143+0.342857\nonumber\\
&=&-0.161113.
\label{43}
\end{eqnarray}
Thus the transition for $k=1$ is of first order. One may note that
roughly $I_2(k=1)\approx I_1(k=1)/{10}$, which is roughly due to a factor
``$12$'' in $g_{v,2}(k=1,x)$. 
The corresponding plot of action $S$ versus $1/T$ is shown in
Fig.5.
Putting all our results together we see that as we go to smaller values
of $k$ the tendency away from second order behavior is even
enhanced, i.e. the transition becomes even more of first order.

\vskip 1cm

{\bf Acknowledgment:} D. K. P. acknowledges support by DAAD and the
Deutsche Forschungsgemeinschaft(DFG), J.--Q. L. acknowledges support
by a DAAD--K.C. Wong Felloship and the Deutsche Forschungsgemeinschaft(DFG).
This work was also supported by Korea Research Foundation(1999-015-DP0074).

\newpage

\begin{appendix}{\centerline{\bf Appendix A: Exact
 calculation of $g_{v,1}(k,x)$}}

\setcounter{equation}{0}
\renewcommand{\theequation}{A.\arabic{equation}}

We set
\begin{equation}
g_{v,1}(k,x)\equiv \frac{k}{2}C^2_1q(k,x)
\label{A.1}
\end{equation}
where
\begin{equation}
q(k,x)={\hat h}^{-1}_v\bigg[\bigg\{(5+4k^2)-9k^2sn^2(x)\bigg\}
sn(x)cn^2(x)dn(x)\bigg].
\label{A.2}
\end{equation}
Hence
\begin{equation}
{\hat h}_v q(k,x)=\bigg[(5+4k^2)-9k^2sn^2(x)\bigg]sn(x)cn^2(x)dn(x).
\label{A.3}
\end{equation}
We multiply the expression by $cn(x)$ and integrate over $x$ from $-K$ to $x$.
Then integrating by parts one obtains
\begin{eqnarray}
&\bigg[&q\frac{d}{dx}cn(x)-cn(x)\frac{dq}{dx}\bigg]^{x}_{x=-K}
+\int^{x}_{-K}dx q.{\hat h}_v cn(x)\nonumber\\
&=&\int^x_{-K}dx \bigg[(5+4k^2)-9k^2 sn^2(x)\bigg]sn(x) cn^3(x) dn(x).
\label{A.4}
\end{eqnarray}
Since $cn(x)$ is the zero mode of ${\hat h}_v$, the second term
on the left hand side of this equation is zero. After integrating
the right hand side of eq.(\ref{A.4}) we obtain
\begin{equation}
\frac{dq}{dx}+\frac{sn(x) dn(x)}{cn(x)}q = \frac{5}{4}(1-k^2)cn^3(x) +
\frac{3}{2}k^2cn^5x +\frac{R}{cn(x)}
\label{A.5}
\end{equation}
where
\begin{equation}
R\equiv\bigg[cn(x)\frac{dq}{dx}+sn(x) dn(x) q(x)\bigg]_{x=-K}.
\label{A.6}
\end{equation}
Eq.(\ref{A.5}) is a simple linear differential equation and its solution is
\begin{eqnarray}
q&=& A cn(x)
+cn(x)\bigg[\frac{1+k^2(2-4R)-3k^4}{4k^2(1-k^2)}E(x)
+\frac{-1+4k^2R+k^4}{4k^2}x\nonumber\\
&+&\frac{1}{2}sn(x) cn(x) dn(x) +\frac{R}{1-k^2}\frac{sn(x) dn(x)}{cn(x)}
\bigg]
\label{A.7}
\end{eqnarray}
where
\begin{equation}
E(x)\equiv\int^{am x}_0\sqrt{1-k^2\sin^2\theta}d\theta =\int^x_0dn^2(u) du.
\label{A.8}
\end{equation}
It is now necessary to determine $A$ and $R$.  
Since we work with partially anti--periodic boundary conditions,
we require
\begin{equation}
q(x+2K)=-q(x).
\label{A.9}
\end{equation}
Using also $E(x+2K)=2E + E(x)$, one can show that
\begin{equation}
R=\frac{(1-k^2)\bigg[(1+3k^2)E-(1-k^2)(1+k^2)K\bigg]}{4k^2\bigg[E-(1-k^2)K\bigg]}
\label{A.10}.
\end{equation}
The expression can also be derived from the definition of $R$ in eq.(\ref{A.6}).
In order to determine $A$ we recall that $cn(x)$ is simply the zero
mode term.  Since, obviously, $q(x)$ does not have a zero mode
component, we
need the condition
$$
\int^K_{-K}dx cn(x) q(x) = 0
$$
from which we conclude that $A=0$.

\end{appendix}

\newpage
\begin{center}
  {\bf Figure Captions}
\end{center}

\noindent

Fig. 1: $I_1(k)$ exact (solid line) and approximate (dashed line)

Fig. 2: $I_2(k)$
 
Fig. 3:  $I_3(k)$
 
Fig. 4:  $I_4(k)$
 
Fig. 5: Action $S$ versus $\tau \equiv 1/T$ demonstrating
the first order transition, the bold line determinig the
transition rate

\begin{figure}
   \begin{center}
      \begin{turn}{-90}
	\includegraphics[height=15cm,keepaspectratio=true]{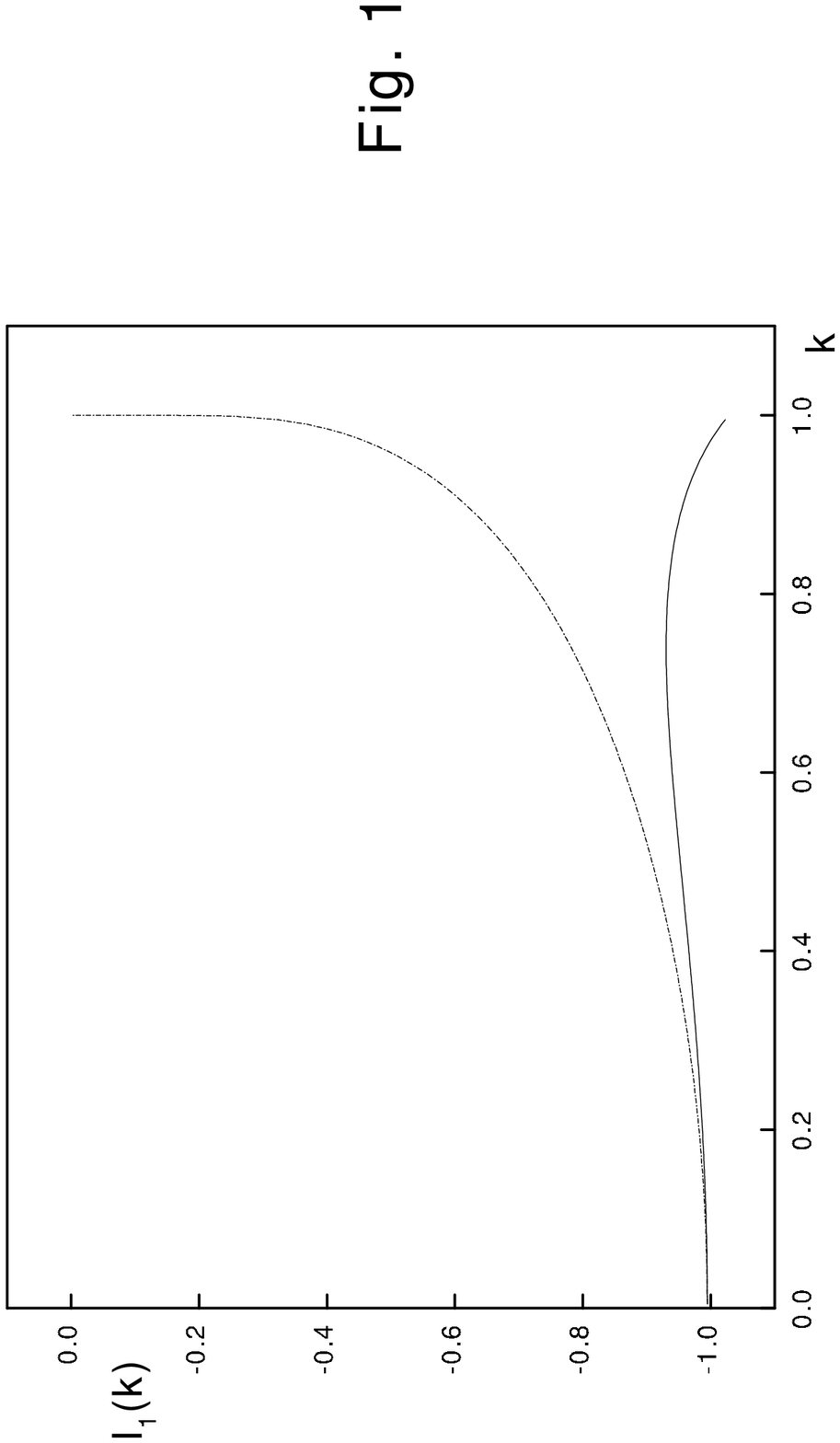}
      \end{turn}
   \end{center}    
   \caption{ $I_1(k)$ exact (solid line) and approximate (dashed line)}
\end{figure}
\begin{figure}
   \begin{center}
      \begin{turn}{-90}
	\includegraphics[height=15cm,keepaspectratio=true]{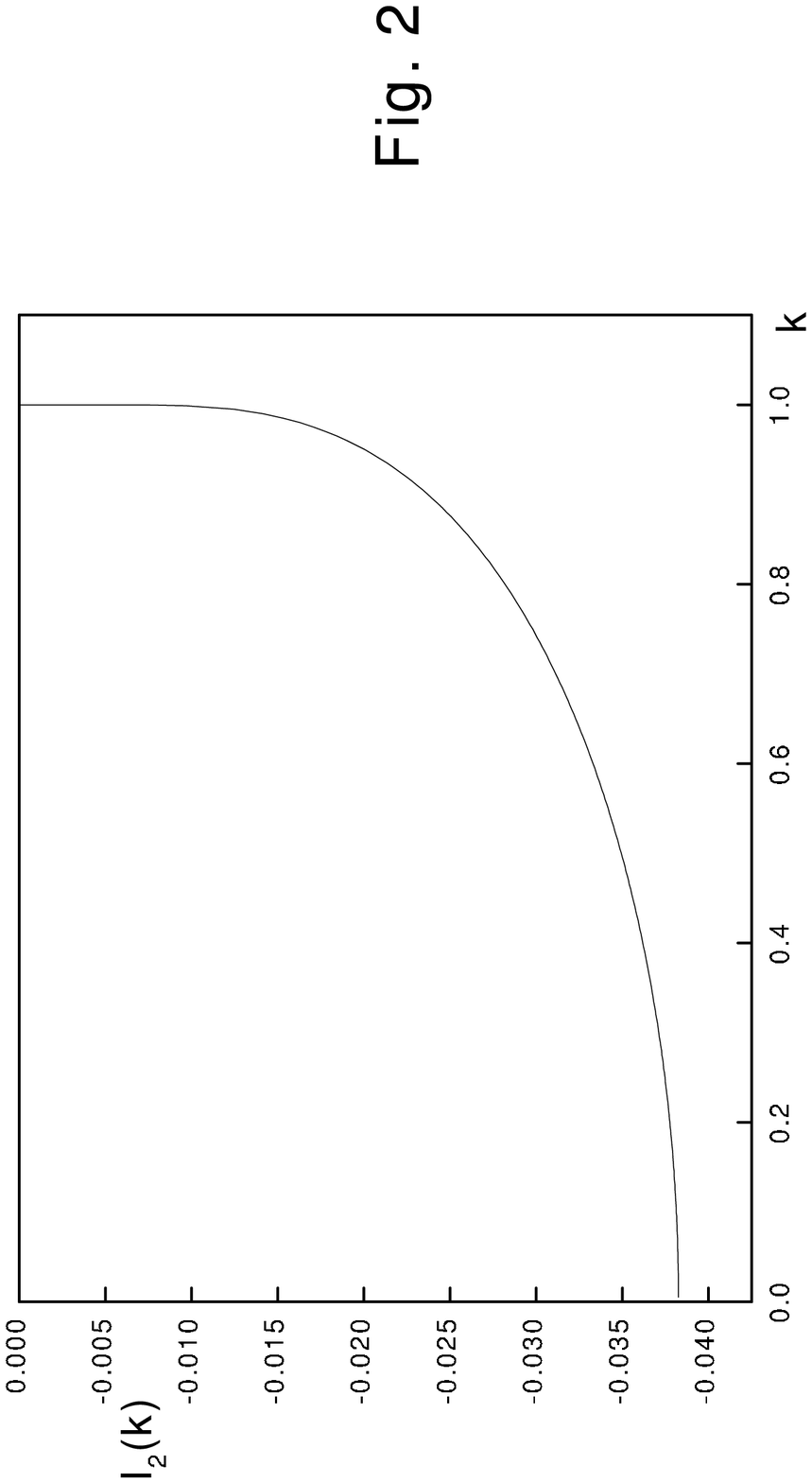}
      \end{turn}
   \end{center}
   \caption{ $I_2(k)$}
\end{figure}
\begin{figure}
   \begin{center}
      \begin{turn}{-90}
	\includegraphics[height=15cm,keepaspectratio=true]{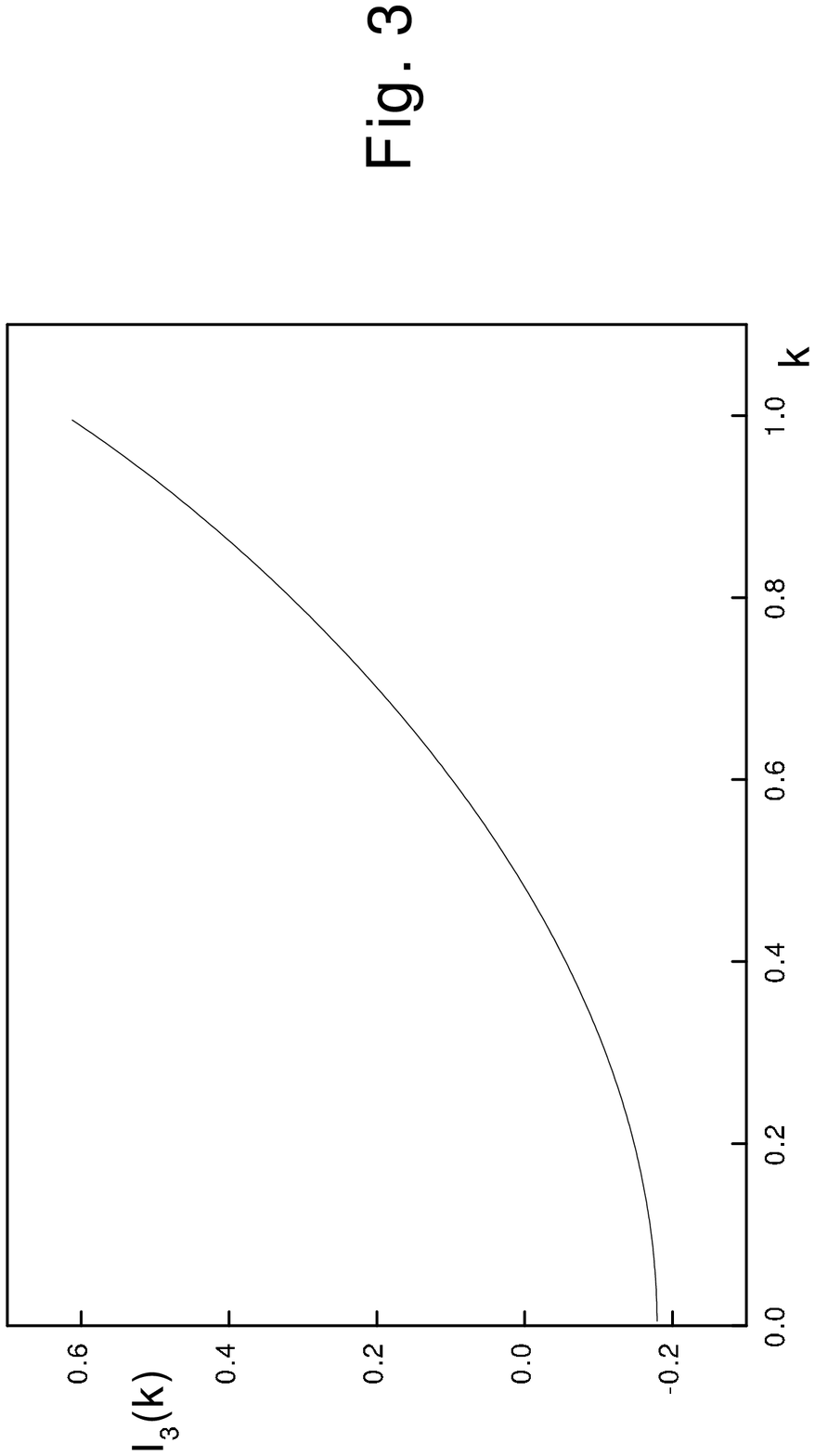}
      \end{turn}
   \end{center}
   \caption{ $I_3(k)$}
\end{figure}
\begin{figure}
   \begin{center}
      \begin{turn}{-90}
	\includegraphics[height=15cm,keepaspectratio=true]{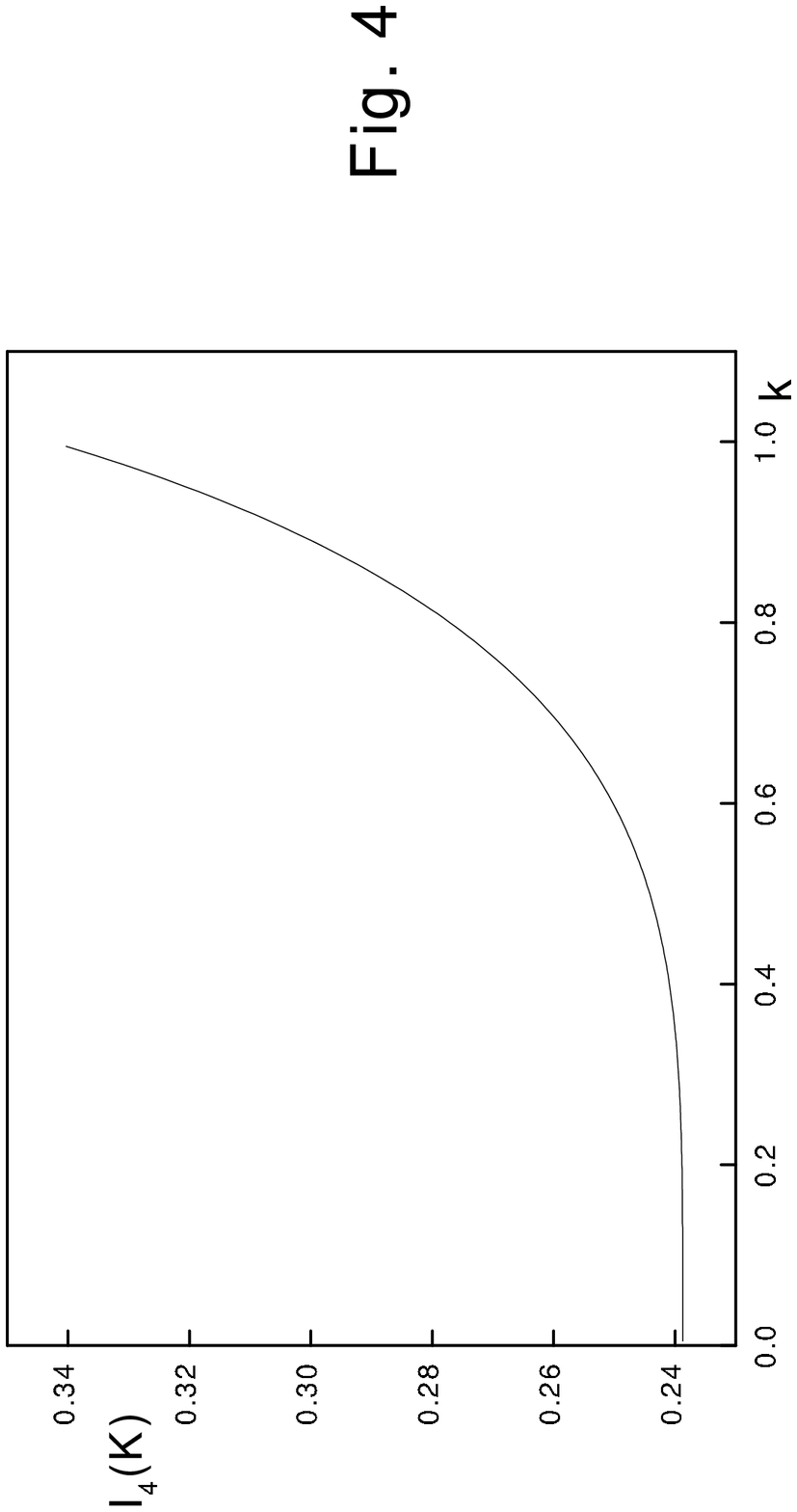}
      \end{turn}
   \end{center}
   \caption{ $I_4(k)$}
\end{figure}
\begin{figure}
   \begin{center}
      \begin{turn}{-90}
	\includegraphics[height=15cm,keepaspectratio=true]{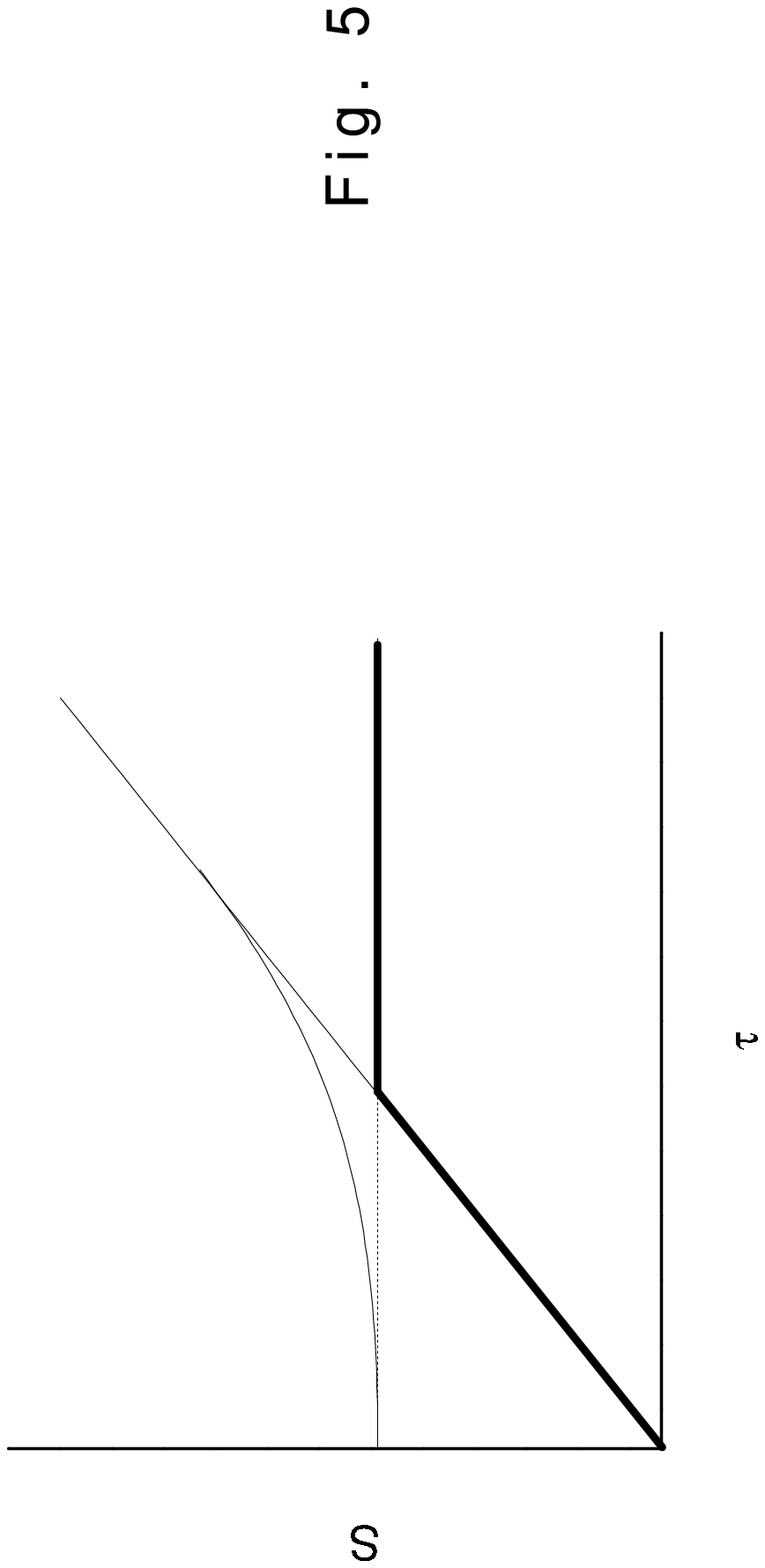}
      \end{turn}
   \end{center}
   \caption{Action $S$ versus $\tau \equiv 1/T$
 demonstrating the first order transition, the bold line
determining the transition rate}
\end{figure}
\end{document}